\documentclass[fleqn,twoside,psfig]{article}
\usepackage{espcrc2}

\usepackage{graphicx}
\usepackage{epsfig}
\usepackage{axodraw}

\newcommand{\be}{\begin{equation}}
\newcommand{\ee}{\end{equation}}
\newcommand{\bea}{\begin{eqnarray}}
\newcommand{\eea}{\end{eqnarray}}
\newcommand{\bi}{\begin{itemize}}
\newcommand{\ei}{\end{itemize}}
\newcommand{\ben}{\begin{enumerate}}
\newcommand{\een}{\end{enumerate}}
\newcommand{\bt}{\begin{tabbing}}
\newcommand{\et}{\end{tabbing}}

\newcommand{\lag}{2{\cal L(C)}}

\newcommand{\AmS}{{\protect\the\textfont2
  A\kern-.1667em\lower.5ex\hbox{M}\kern-.125emS}}

\title{
\vspace*{-2.cm}
\begin{flushright}
{\normalsize UTHEP-496}\\ 
{\normalsize UTCCS-P-7}\\
\end{flushright}
A scaling study of the step scaling function 
of quenched QCD
\\
with improved gauge actions
\thanks{Talk presented by S.\ Takeda
This work is supported in part by Grants-in-Aid for
Scientific Research from the Ministry of Education, 
Culture, Sports, Science and Technology.
(Nos. 13135204, 
13640260, 
14046202, 
14740173, 
15204015, 
15540251, 
15540279, 
15740134, 
16028201, 
16540228, 
16740147, 
16$\cdot$11968). 
S.T. is supported by the JSPS Research Fellowship.
}}

\author{
S. Takeda\address[TSUKUBA]
{Grad. School of Pure and Applied Sciences, University of Tsukuba, 
Tsukuba, Ibaraki 305-8571, Japan},
S. Aoki\addressmark[TSUKUBA], 
M. Fukugita\address[ICRR]
{Institute for Cosmic Ray Research, University of Tokyo, 
Kashiwa 277-8582, Japan},
K-I. Ishikawa\address[HIROSHIMA]
{Department of Physics, Hiroshima University, 
Higashi-Hiroshima, Hiroshima 739-8526, Japan}, 
N. Ishizuka\addressmark[TSUKUBA]$^{,}$\address[CCS]
{Center for Computational Sciences, University of Tsukuba, 
Tsukuba, Ibaraki 305-8577, Japan},
Y. Iwasaki\addressmark[TSUKUBA],\\
K. Kanaya\addressmark[TSUKUBA], 
T. Kaneko\address[KEK]
{High Energy Accelerator Research Organization (KEK), 
Tsukuba, Ibaraki 305-0801, Japan},
Y. Kuramashi\addressmark[TSUKUBA]$^{,}$\addressmark[CCS], 
M. Okawa\addressmark[HIROSHIMA],
Y. Taniguchi\addressmark[TSUKUBA]$^{,}$\addressmark[CCS], 
A. Ukawa\addressmark[TSUKUBA]$^{,}$\addressmark[CCS],
T. Yoshi\'e\addressmark[TSUKUBA]$^{,}$\addressmark[CCS] \\
 (CP-PACS Collaboration)
}

\begin{document}
\pagestyle{empty}

\begin{abstract}
We study the scaling behavior of the step scaling function for 
SU(3) gauge theory, employing the Iwasaki gauge action 
and the Luescher-Weisz gauge action.
In particular,
we test the choice of boundary counter terms and apply
a perturbative procedure for
removal of lattice artifacts for the simulation results
in the extrapolation procedure.
We confirm the universality of the step scaling functions
at both weak and strong coupling regions. 
We also measure the low energy scale ratio with the Iwasaki action,
and confirm its universality.
\end{abstract}

\maketitle

\section{Introduction}
\label{sec:Introduction}

Recently CP-PACS and JLQCD Collaborations
have started a project for $N_{\rm f}=3$ QCD simulations.
One of the targets of the project is to evaluate the strong coupling constant 
$\alpha_{\overline{\rm{MS}}}$ in the $N_{\rm{f}}=3$ QCD  
using the Schr\"odinger functional(SF) scheme\cite{sforiginal}.
In the project Iwasaki gauge action\cite{Iwasaki} has been
employed to avoid the strong lattice artifacts of the plaquette gauge
action found in $N_{\rm f}=3$ simulations\cite{okawa}.

In a previous study \cite{takeda}, as our first step toward evaluation of 
$\alpha_{\overline{\rm{MS}}}$ for $N_{\rm{f}}=3$, 
O($a$) boundary improvement coefficients in the SF scheme
have been determined for the various improved gauge actions
up to one-loop order in the perturbation theory.
As the next step, 
we investigate the lattice cut off dependence of 
the step scaling function(SSF)
non-perturbatively in quenched lattice QCD simulations with
the improved gauge actions
for some choices of the improvement coefficients. 
We also confirm the universality of SSF and the low
energy scale ratio, by comparing our results with the previous ones obtained
by ALPHA Collaboration\cite{su3,NPRquench,Necco}.
We refer to ref. \cite{SFRG} for unexplained notations
in this report.

\section{Setup}
\label{sec:Setup}
\begin{figure*} [tbh]
\vspace{-2mm}
\begin{center} 
\begin{tabular}{cc}
\resizebox{73mm}{!}{\includegraphics{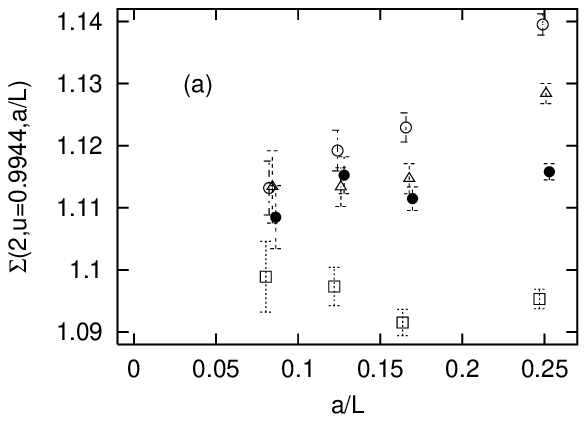}}
\resizebox{73mm}{!}{\includegraphics{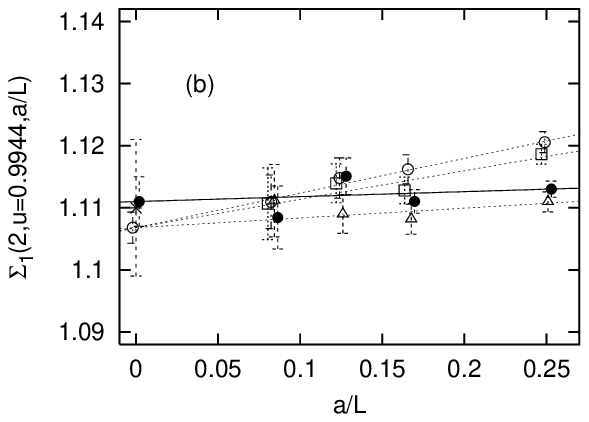}}
\\
\resizebox{73mm}{!}{\includegraphics{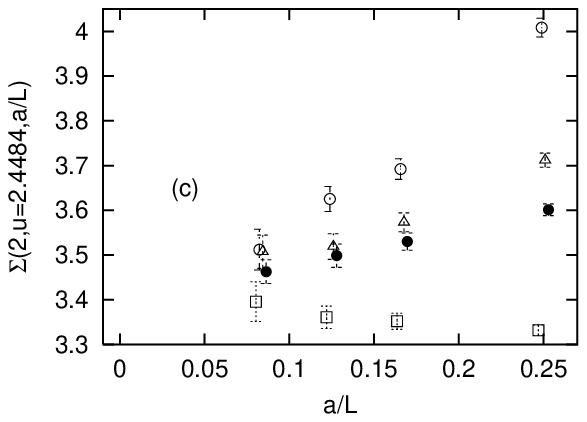}}
\resizebox{73mm}{!}{\includegraphics{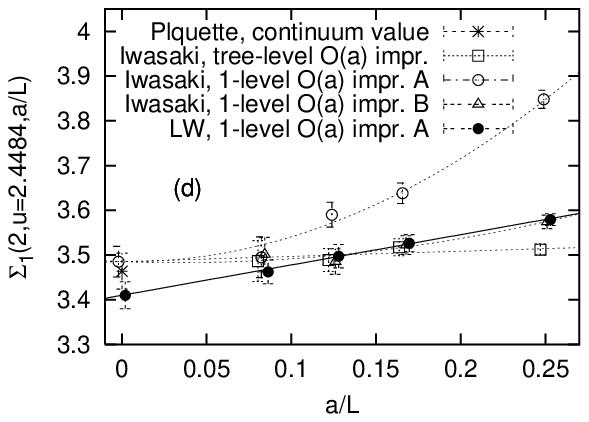}}
\vspace{-8mm}
\end{tabular}
\end{center} 
\caption{The cut off dependence of 
$\Sigma$(left-hand side)
and $\Sigma_1$(right-hand side)
at the weak $u=0.9944$(upper) and the strong $u=2.4484$(bottom) couplings 
for the various improved gauge actions.
The points at $a/L=0$ in (b) and (d) represent continuum values
for the Iwasaki action(open circle),
the LW action(fulled circle)
and also the plaquette action(cross)\cite{su3,NPRquench}.
The dotted lines and the straight lines
represent the combined fit function for the Iwasaki action
and the liner fit function for the LW action respectively.
} 
\label{fig:SSF}
\end{figure*}

We follow the same SF setup 
as in the case of the Wilson plaquette action\cite{su3},
but employ the improved gauge
actions including the plaquette(${\cal S}_0$) and rectangle(${\cal S}_1$)
 loops:
    \begin{equation}
      S_{\rm{imp}}[U] 
      =
      \frac{1}{g^2_0} 
      \sum_{i=0}^{1}
      \sum_{{\cal C} \in {\cal S}_i} W_i({\cal C},g^2_0) \lag.
     \label{eqn:impaction}
    \end{equation}
The assignment of the weight factor $W_i({\cal C},g^2_0)$
near the boundary in the time direction
is important to achieve O($a$) improvement in the SF scheme.
We employ the weights,
\be
      c_0 c^P_{\rm{t}}(g^2_0)
      = 
      c_0 ( 1 + c^{P(1)}_{\rm{t}} g^2_0 + O(g^4_0) ), 
\ee
\be
      c_1 c^R_{\rm{t}}(g^2_0)
      = 
      c_1 ( 3/2 + c^{R(1)}_{\rm{t}} g^2_0 + O(g^4_0) ),
\ee
where $c_0+8c_1=1$,
for the plaquette 
and rectangular loops(having 2-link for the spatial direction)
touching the boundary respectively.
The leading terms necessary for the tree-level O($a$) improvement
is taken from Ref.\cite{aokiweisz}.

The one-loop terms\cite{takeda}
have to satisfy the following relation
to achieve one-loop O($a$) improvement:
$c_0 c^{P(1)}_{\rm{t}} + 4 c_1 c^{R(1)}_{\rm{t}}=A_1/2$
where $A_1$ is a coefficient of $g^4_0 a/L$ term in the SF coupling.
We consider two choices: 
$c^{R(1)}_{\rm{t}}=2 c^{P(1)}_{\rm{t}}$ (condition A),
and $c^{R(1)}_{\rm{t}}=0$ (condition B).
While the difference between the conditions A and B
is an O($a^5$) contribution at one-loop level,
it may become larger at higher-loop orders.

Let us introduce quantities which we calculate.
The SSF describes the evolution
of the running coupling under a finite scaling factor(say $s=2$).
$\Sigma$
denotes the SSF calculated on the lattice.
In a step to fix the energy scale, 
one have to set
the low energy scale ratio
$L_{\max}/r_0$
at the continuum limit
where $L_{\max}$ is defined 
as $\bar{g}^2(L_{\max})=3.480$
and $r_0$ is the Sommer's scale.
Later on,
we shall show the simulation results
of their cut off dependence
and take the continuum limit.

\section{Results}
\label{sec:Results}

Simulations for SSF and $r_0/a$ 
are performed on CP-PACS.
We refer to Ref.\cite{SFRG} for details of simulation parameters and 
the continuum extrapolation procedure.

In Fig.~\ref{fig:SSF},
we show the cut off dependence of 
the lattice SSF 
for the improved gauge actions
with some choices of the boundary counter terms.
The panels (a) (for weak coupling $u=0.9944$) 
and (c) (for strong coupling $u=2.4484$) 
show the raw data for $\Sigma$,
while a perturbatively improved observable $\Sigma_1$\cite{su2poly}
is plotted in (b) (for weak) and (d) (for strong).
The latter is defined as
\begin{equation}
  \Sigma_1(2,u,a/L)
  =
  \Sigma(2,u,a/L)/(1 + \delta_1(a/L) u),
\end{equation}
where $\delta_1(a/L)$ is the one-loop relative deviation\cite{takeda}.
We observe that $\Sigma_1$
shows a better scaling behavior at both coupling regions 
than $\Sigma$.
It is particuraly so for the case of the Iwasaki action with
the tree level O($a$) improvement and the condition B.
Therefore,
we use $\Sigma_1$
in the continuum extrapolation.
For the Iwasaki action
the continuum value is obtained by a joint fit of the three sets of data.
For the LW action,
we take the value obtained by a linear fitting 
of the data with the condition A.
The continuum values are plotted at $a/L=0$ in Fig. \ref{fig:SSF} (b) and (d).
We observe
that the three values obtained 
with the Iwasaki and LW actions in the present work 
and that of ALPHA Collaboration \cite{NPRquench}
are consistent within 
1$\sigma$ ($2.3 \sigma$) 
at the weak (strong) coupling.  

\begin{figure}[t!]
\resizebox{75mm}{50mm}{\includegraphics{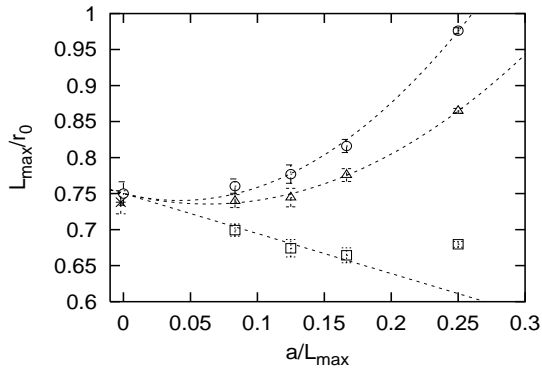}}
\vspace{-10mm}
\caption{The cut off dependence of the low energy scale ratio
for the Iwasaki action.
The symbols are the same as the case of the SSF.
} 
\vspace{-4mm}
\label{fig:lowenergyscale}
\end{figure}

We also show
the low energy scale ratio for the Iwasaki action
with both tree-level and one-loop O($a$) improvements
in Fig.~\ref{fig:lowenergyscale}.
To extrapolate to the continuum limit,
we use the same fitting form as in the case of SSF at strong coupling. 
We do not include the point $L/a=4$ for the tree level O($a$) improved case.
We observe that
the extrapolated value
agrees with the one obtained by the 
plaquette action\cite{Necco} within errors.

\section{Conclusions}
\label{sec:Conclusions}

We have confirmed the universality of
the SSF and the low energy scale ratio.
In the extrapolation procedure,
the perturbative removal of lattice artifacts well reduces 
the scaling violation of the SSF
for the Iwasaki action
with the tree level O($a$) improvement and the
one-loop O($a$) improvement with the condition B.

As mentioned in the introduction,
this work is the second step
toward unquenched simulations.
The present study shows that we should use
the above two choices
in the future simulations with dynamical quarks.


%


\end{document}